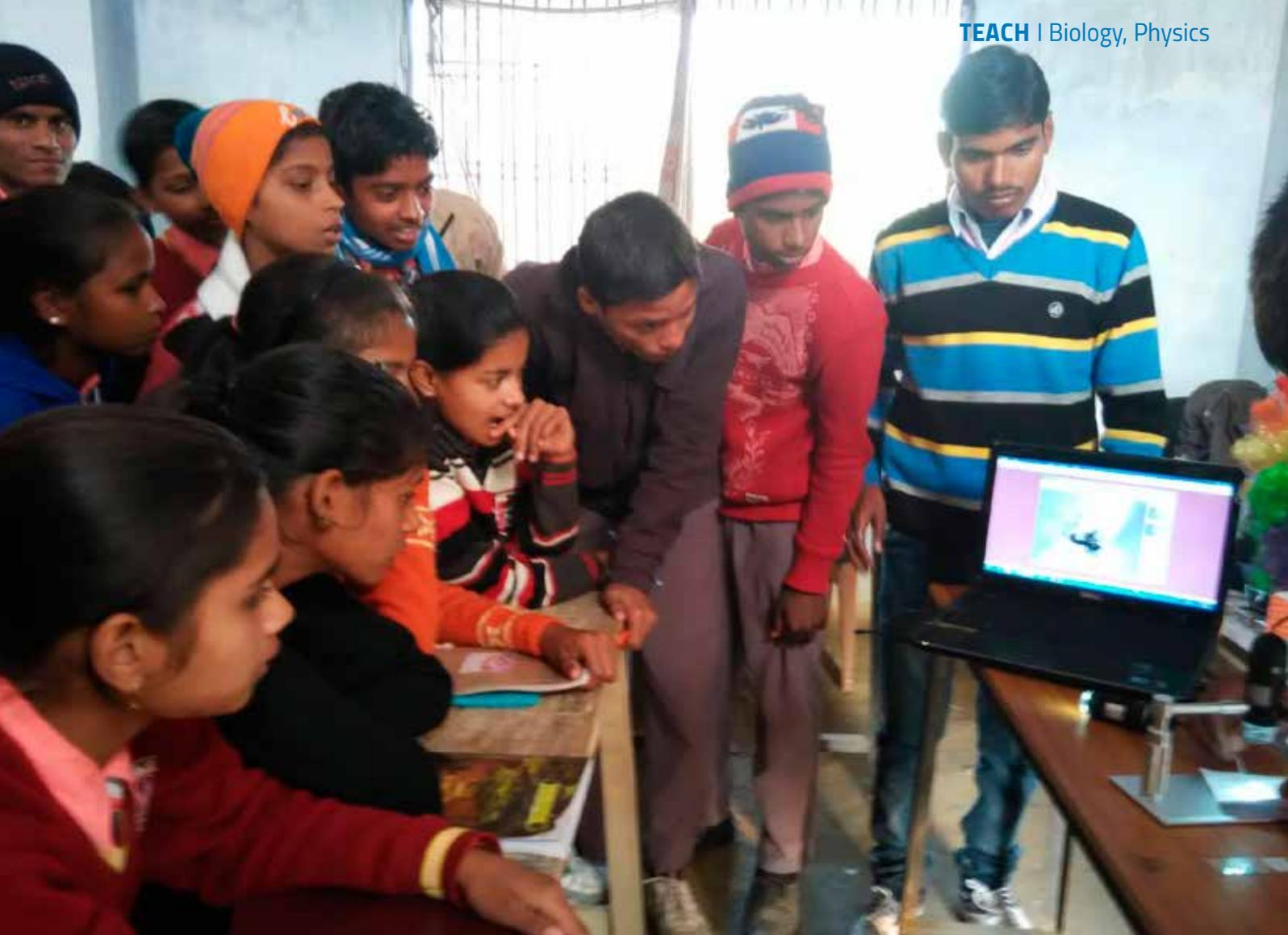

Image courtesy of Anand Singh and Tim Saunders

# Doing is understanding: science fun in India

School children in India built their own digital microscope, bent light and investigated gas laws. Find out how.

By Anand Pratap Singh, Anuradha Gupta, Ranjit Gulvady, Amit Mhamane and Timothy Edward Saunders

In India, as in many countries, the main focus in science classrooms is on exams rather than musing on the fascinating concepts and understanding of the world that science offers. This can mean that students lose interest in studying science – a problem that is further hampered where there is a lack of facilities, expertise or mentors. We started the 'Science is fun' outreach programme to address these problems. The 15-person team, led by undergraduate and research scientists, conducted four workshops with underprivileged children in Indian primary and secondary schools during December 2014 and January 2015.

The workshops explored basic science concepts, reinforced by hands-on experiments using readily available materials. They were generally successful, with students keen to participate and motivated to learn more after the workshops. We were also pleasantly surprised to see students engaging with new concepts and not hesitating to participate in the discussions. We tried to ensure teachers were central to the activities, and also designed the experiments to be easily repeatable so that teachers could incorporate them into their own lessons once the workshops were over.

In this article, we describe three of our successful activities: building a periscope and a digital microscope, and two experiments based on the physical gas laws. All are cheap and easy to perform, yet reveal interesting scientific principles. Each activity takes about an hour.

## Periscopes and the reflection of light

Periscopes allow us to look over or around obstacles without being in the line of sight, and they are still used today in submarines. The most basic version involves two plane mirrors, one placed at each end of a long tube, with the mirrors parallel to each other and at 45° to the length of the tube.

This simple periscope is easy to build in around one hour and requires no more materials than cardboard, mirrors, scissors and sticky tape. It can be used with younger children who can safely handle scissors and mirrors (e.g. 8–13),

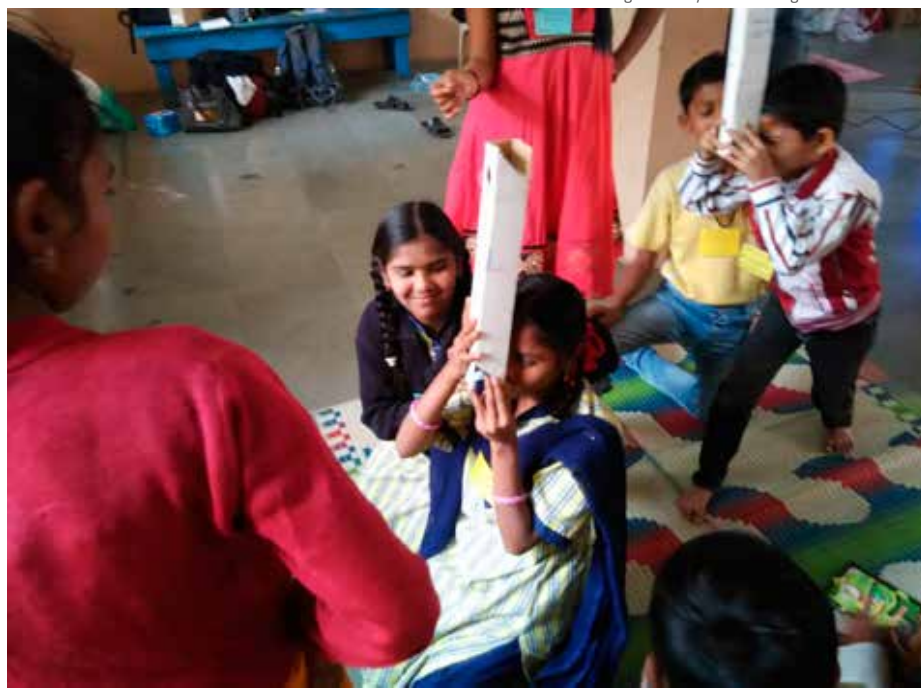

*Students using the simple periscope*

as a fun activity and an introduction to the reflection of light. Older students (e.g. 14–16) could use it to learn about the reflection of plane mirrors. Figure 1 shows how the periscope is made and full assembly instructions can be downloaded from the *Science in School* website[w1].

Although the basic design of the periscope is simple and straightforward, it demonstrates the law of reflection: the angle of incidence of a light ray is equal to its angle of reflection (figure 1). Since the angles of incidence and reflection are the same, light that is travelling horizontally to hit a mirror orientated at 45° will then travel vertically. By adding another mirror angled at 45°, the light path is re-orientated to travel horizontally again, but now it is displaced from the original beam path by the height of the periscope.

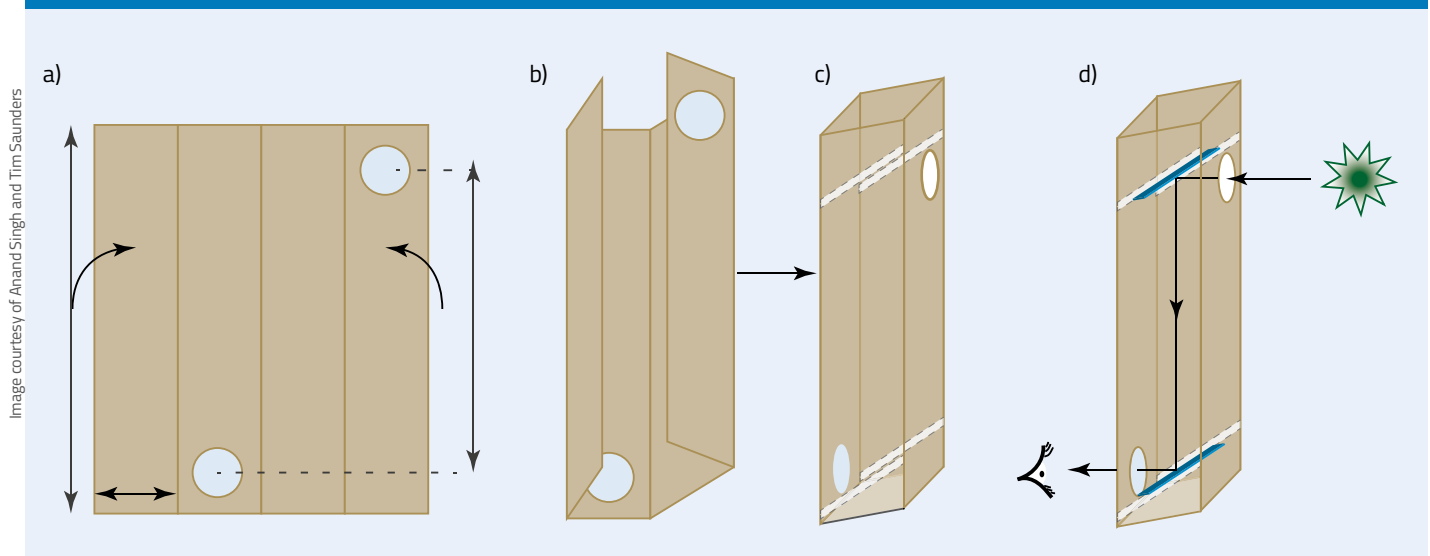

*Figure 1: A periscope can be simply built from a piece of cardboard and two mirrors. a) Cut two holes in the cardboard. b) Fold the cardboard and tape it closed. c) Cut slits for the mirrors. d) Insert the mirrors and your periscope is ready to use.*

By building their own periscope, students can see for themselves that light travels in straight lines and that the direction of travel can be altered by reflective surfaces: a common use of periscopes in optics is to change the height of a light path. This activity can be extended to be more challenging for older students by requiring a different angle of observation[w2].

Suitable questions for your students could include:

- Periscopes have been used extensively in warfare. What examples can you think of and what are the advantages and disadvantages of periscopes for the military?

  Periscopes are / were used in submarines, tanks and in trench warfare. Although they allow the viewer to make observations from a position of safety, they are only two-dimensional in that they cannot look up or down without additional components.

- What role does science have in military conflicts? Should scientists work to develop better equipment for the military?

  See also Essex & Howes (2014) for a discussion of the chemical legacy of war.

- To use a submarine periscope, the viewer turns the whole periscope around, walking around the main tube. This requires a lot of space – which is limited inside submarines. Why, then, do the periscopes not simply rotate at the top?

- How would you adapt a periscope to be able to look up and down?

## Building and using a digital microscope

Microscopes allow us to explore details that are too small to be seen by the naked eye, but they are usually expensive. Here, we show how to build a cheap webcam-based microscope, so that students can learn the basics of optics and image formation while exploring the microscopic world around us.

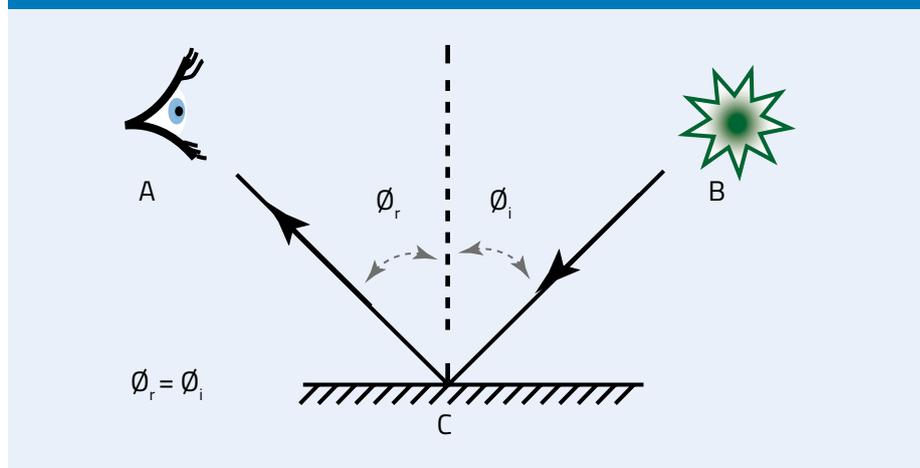

*Figure 2: The reflection of light. A: observer's eye; B: object; C: mirror; Øi: incident angle; Ør: reflected angle*

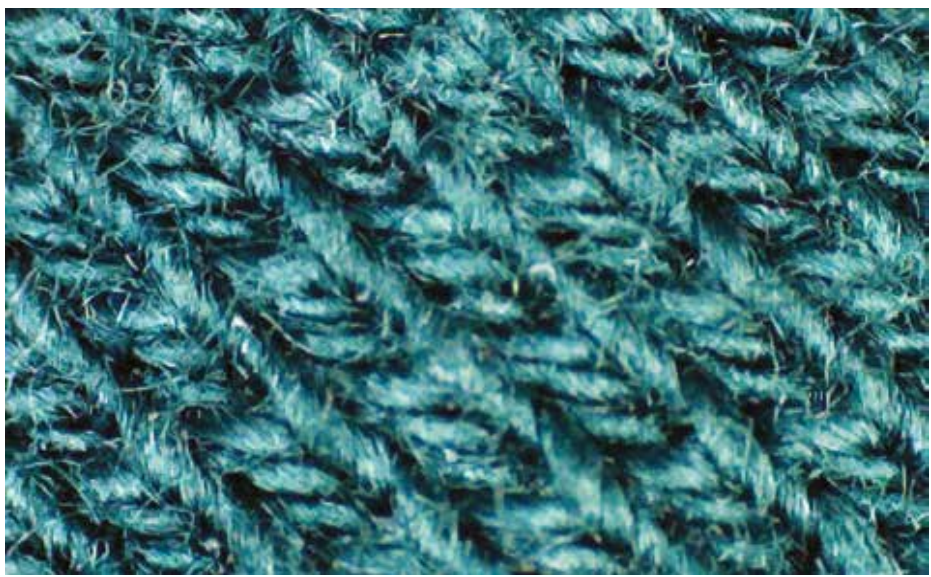

*Fabric observed under the digital microscope*

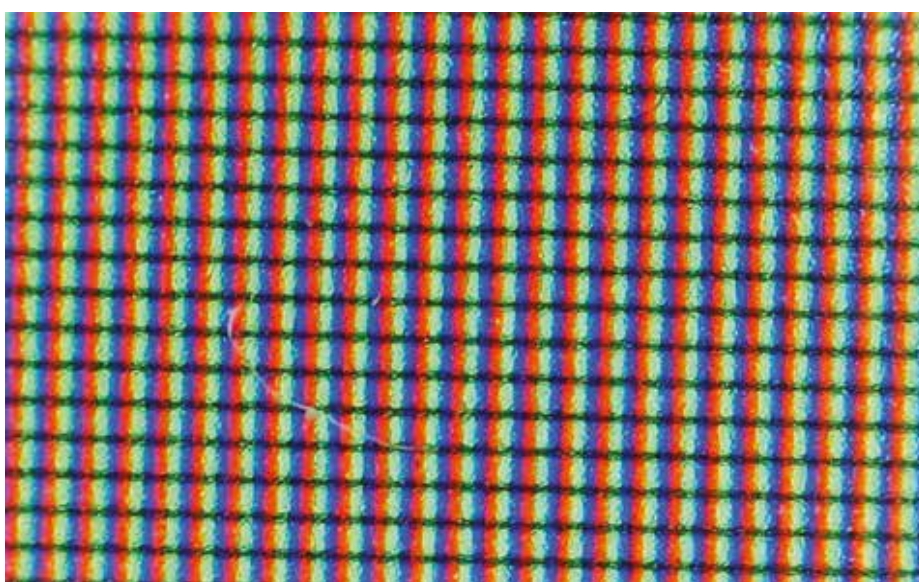

*Pixels on the computer screen, as seen with the digital microscope*

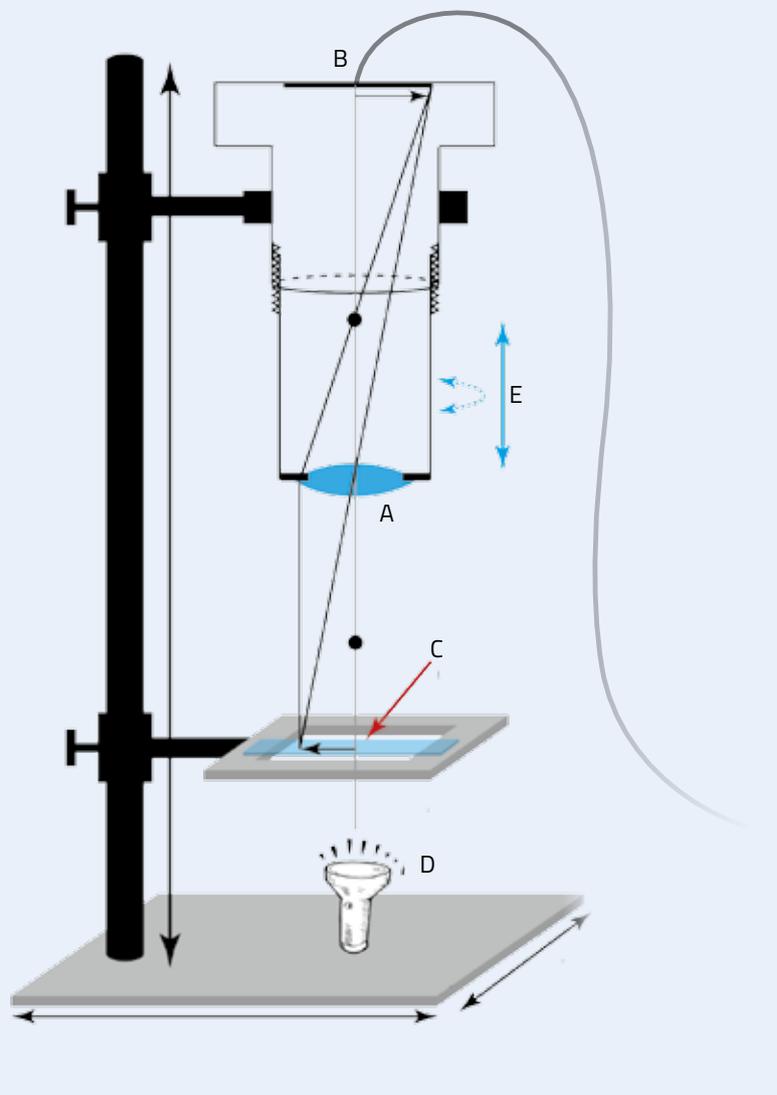

*Figure 3: Supported by a clamp stand, the homemade microscope consists of a plastic tube with a lens (A) attached to one end and a webcam (B) to the other, attached to a computer. The sample (C) is clamped in place below the lens and a torch (D) is shone on it. The image can be zoomed by lengthening or shortening the tube (E).*

Image courtesy of Anand Singh and Tim Saunders

unique view into the lives of plants and animals.

Constructing the microscope takes around one hour and can be done by students aged 14 years and over, perhaps in a small group. Allow another hour for using the microscope, an activity that is suitable for students of all ages.

## Using your microscope

To start with, you could ask your students to examine the following samples under the microscope:

- Sugar, salt and sand grains
- Feather, hair, fabric or the torn edge of a piece of newspaper
- Mould on bread, fruit or vegetables
- Pond water to see microscopic organisms.

Using image software (e.g. the free software ImageJ[w3]), they could then use quantitative measurements to make some forensic investigations, such as:

- Distinguishing hairs from different people based on their thickness
- Distinguishing hairs from different animal species based on colour, thickness or other characteristics
- Distinguishing sand or soil samples from different locations
- Distinguishing grains of sugar from grains of salt.

The microscope consists of a plastic tube with a lens attached to one end and a webcam to the other, attached to a computer. The microscope is supported by a clamp stand, the sample to be viewed is clamped in place below the lens, and a torch is shone onto or through the sample. The image is focused by the lens onto the image sensor of the webcam and thus passed to the computer for analysis. By adjusting the length of the tube or the relative positions of the microscope and sample, the image can be zoomed and focused. Figure 3 gives an overview of the construction, and full instructions for building and adjusting the microscope can be downloaded from the *Science in School* website[w1].

The key difference between a digital microscope and a standard (analogue) school microscope is that students can take digital pictures that can be stored and analysed in later lessons. They can also easily make quantitative measurements using appropriate software. Furthermore, if resources are limited, the output from a single digital microscope can be projected onto a screen for the whole class to see. Finally, the digital microscope can be used to make videos, allowing a

The digital microscope could also be used to quantify variability in microscopic samples. For example, your students could take 50 images of hair and measure the widths to find the mean and standard deviation of the hair width. Do they differ between hair colours? Between animal species? Such investigations are much more difficult to do with an analogue microscope.

By making movies with their digital microscope, your students could record the tracks of pond animals and compare the patterns of movement of different species. Over a longer period of time, they could even watch plants grow. For example, they could leave a plant root under the microscope for a week and get the computer to record an image every hour. Using the resulting movie, they could then plot a growth curve.

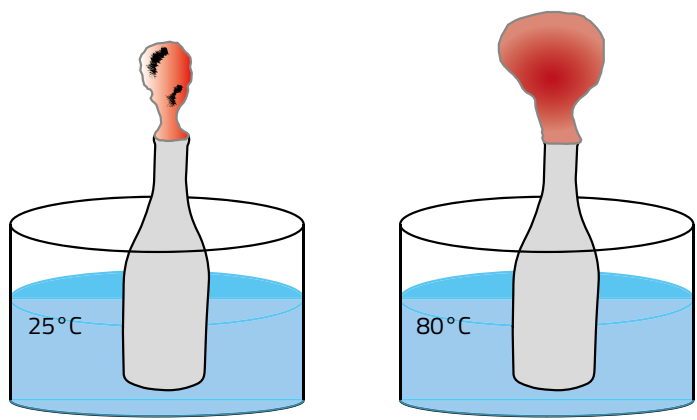

*Figure 4: Demonstrating Charles' gas law. When placed in hot water, the air in the bottle expands, filling the balloon.*

## Gas laws

### Demonstrating Charles' law

We all enjoy playing with balloons or watching hot-air balloons, but these activities are rarely related to the gas laws learned in the classroom. We therefore developed an experiment to show the relationship between the temperature and volume of a gas. This relationship is described by Charles' law, which states that, for a fixed amount of gas at constant pressure, the volume of the gas is directly proportional to its temperature.

This experiment takes around one hour and is suitable for students from around 14 years old. For older students, the activity can be adapted to experimentally estimate absolute zero.

### Materials
- Balloons
- Hot (approximately 80°C) and room-temperature water
- Two dry and empty plastic bottles
- Two water tubs, large enough to fit the empty bottles

### Procedure
1. Fill one of the water tubs with hot water and one with water at room temperature.
2. Place a balloon tightly over the mouth of each plastic bottle.
3. Submerge one bottle in the hot water and the other in the water at room temperature (figure 4).

**Safety note:** Care must be taken when using hot water.

This simple set-up maintains a fixed amount of gas (the contents of the bottle) at a fixed pressure (atmospheric). Once the bottle is placed in hot water, the temperature of the gas inside it increases and the gas expands – as described by Charles' law – and it begins to fill the balloon, as students will see (figure 4). Meanwhile, the volume of the balloon at the room temperature should remain the same.

This process is reversible. Remove the bottle from the hot water and place it in the water at room temperature: the balloon will slowly shrink again.

Another simple demonstration of Charles' law is to inflate a balloon, then place it in a freezer for about an hour and observe the change in volume. We also used balloons to demonstrate Boyle's law[w1] – which states that at a constant temperature, the pressure of a fixed amount of gas changes inversely with its volume.

### Estimating absolute zero

For older students (ages 16–19), absolute temperature can be roughly estimated by measuring the gas volume of a frictionless syringe at different temperatures.

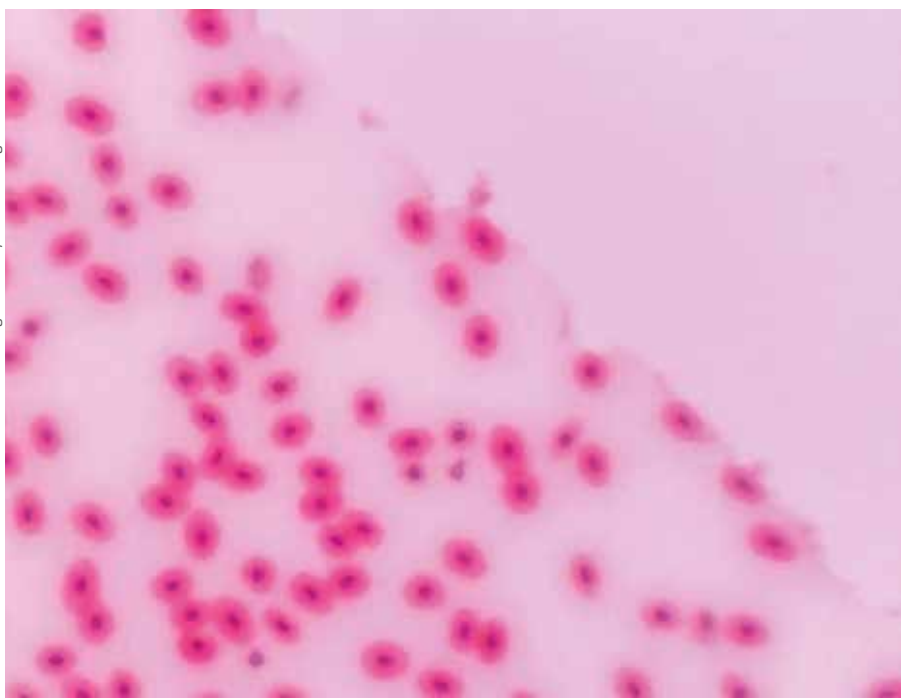

*Images courtesy of Anand Singh and Tim Saunders*

*Frog blood cells observed under the digital microscope*

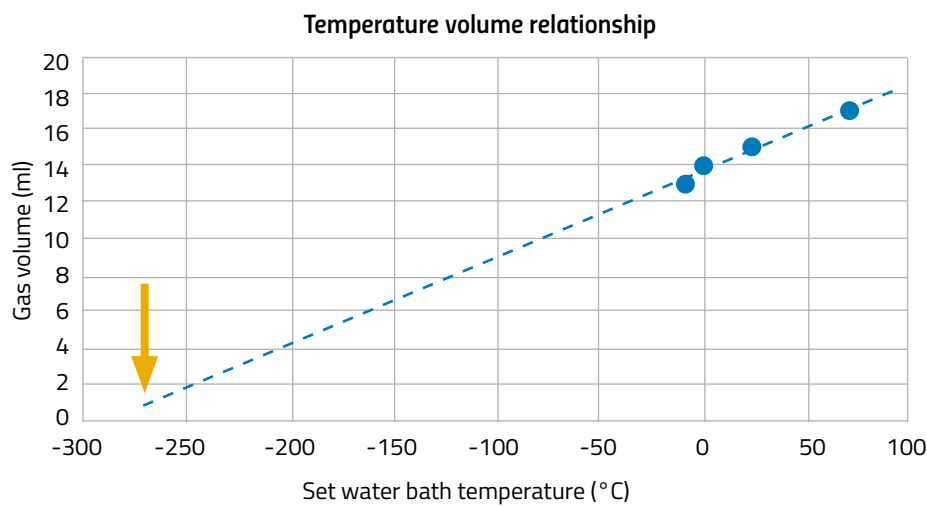

*Figure 5: From a plot of temperature (°C) against gas volume (ml), absolute zero can be estimated by extrapolating to zero volume.*

**Materials**
- Plastic syringe
- Oil or grease
- Modelling clay, glue or syringe caps to seal the syringe tip
- Waterbaths at 0, 10, 22 (room temperature), 50 and 80 °C.

**Procedure**
1. Apply oil or grease to a plastic syringe to make it close to frictionless.
2. Draw a known volume of air into the syringe and seal the tip thoroughly.
3. Measure the air volume at different temperatures, e.g. using water baths at 0, 10, 22 (room temperature), 50 and 80 °C.
4. Plot the temperature (°C, *x* axis) against gas volume (ml, *y* axis).

   You should find a linear relationship between temperature and volume. By extrapolating to zero volume, the corresponding temperature on the *x* axis can be seen to be approximately 270 °C (the actual value of absolute zero is -273.15 °C).

## Web references

w1 Further information about these and other activities used in the workshops can be downloaded from the *Science in School* website. See: www.scienceinschool.org/2015/issue34/india

w2 To learn more about the reflective properties of light, see *Mirror, Mirror on the Wall: Angles of Reflection*. See www.optics4kids.org or use the direct link: http://tinyurl.com/mirrorwall

w3 The imaging software ImageJ can be downloaded free of charge from the website of the US National Institutes of Health. See: http://imagej.nih.gov/ij

w4 Watch a video showing the experiment to calculate absolute zero using a fixed volume of gas at different temperatures. See: https://www.youtube.com/watch?v=wkWo-8tY8cY

## Resources

For visual demonstrations of how lenses, mirrors and other optical components change the light path, see:

A plane mirror reflection experiment. See www.education.com or use the direct link: http://tinyurl.com/o2mf8mc

A video on light experiments using a fish tank: https://www.youtube.com/watch?v=aN1saYRr6z4

An experiment to deflect a laser beam on the EU Hands-on Universe website. See: http://www.pl.euhou.net/docupload/files/Excersises/WorldAroundUs/Refraction/refraction.pdf or use the direct link: http://tinyurl.com/qx3gffu

To learn about early microscopes and how to build a compound microscope using cheap materials, see:

Tsagliotis N (2012) Build your own microscope: following in Robert Hooke's footsteps. *Science in School* **22**: 29-35. www.scienceinschool.org/2012/issue22/microscope

Instructions and supporting material for building and using Galileo's microscope on the website of the Museo Galileo. See: http://www.museogalileo.it or use the direct link: http://tinyurl.com/q244phr

Instructions for building a 'one-dollar microscope' on the Funsci website. See www.funsci.com or use the direct link: http://tinyurl.com/9wvgp

For other microscopy resources, see:

Microworlds, an exploration of the microscopic world on the Funsci website. See www.funsci.com or use the direct link: http://tinyurl.com/oxnrh68

A paper microscope: Cybulski JS, Clements J, Prakash M (2014) Foldscope: origami-based paper microscope. *Plos One* **9(6)**: e98781. doi: 10.1371/journal.pone.0098781
*Plos One* is an open-access journal, so all articles are freely available online.

A microscope on a mobile phone: Smith ZJ et al. (2011) Cell-phone-based platform for biomedical device development and education applications. PLoS ONE, 2011. **6**(3): p. e17150. doi: 10.1371/journal.pone.0017150
*Plos One* is an open-access journal, so all articles are freely available online.

Instructions for preparing slides of garlic or onion cells on the Funsci website. See www.funsci.com or use the direct link: http://tinyurl.com/pajppg6

For instructions for experiments on Charles' law, see the common gas law demonstrations on the website of North Carolina State University. See http://ncsu.edu or use the direct link: http://tinyurl.com/pq7qu6h

For video experiments with balloons on water and air compressibility, see: https://www.youtube.com/watch?v=jJToILG7GAI

For background information relevant to the development of our workshops, see:

Denofrio LA et al (2007) Linking student interests to science curricula. *Science* **318**: 1872–1873. doi: 10.1126/science.1150788

Donald B (2013) Stanford scholars find varying quality of science and tech education in Brazil, Russia, India and China. *Stanford Report*: 19 August. http://news.stanford.edu or use the direct link: http://tinyurl.com/pbmrnvu

Leshner AI (2007) Outreach training needed. *Science*: **315**: 161. doi: 10.1126/science.1138712

## References

Saunders T (2011) The physics of crowds. *Science in School* **21**: 23–27. www.scienceinschool.org/2011/issue21/crowding

Essex J, Howes L (2014) Experiments in integrity – Fritz Haber and the ethics of chemistry. *Science in School* **29**: 5–8. www.scienceinschool.org/2014/issue29/ethical_chemistry

---

Anand Singh is a post-doctoral fellow at the Mechanobiology Institute, National University of Singapore, Singapore (MBI). He loves building new microscopes, and he uses these to study biological processes ranging in scale from the sub-cellular to whole embryos. He enjoys outreach and has been involved in developing simple tools for science workshops and public outreach.

Anuradha Gupta is a post-doctoral fellow at the Inter-University Centre for Astronomy and Astrophysics, Pune, India. She is an astrophysicist and promotes basic education for underprivileged children, especially for girls in India.

Ranjit Gulvady is a fourth-year graduate student at MBI. He is a single-molecule biophysicist who uses light microscopy to understand physical properties of biological molecules like DNA. During high school, he organised several science demonstrations for school children and later in college he helped children from especially underprivileged backgrounds to understand basic science and mathematics.

Amit Mhamane did his bachelor's and master's degrees in biotechnology at the University of Mumbai, India, and now is doing a PhD in chemistry at the Max Planck Institute of Molecular Physiology, Dortmund, Germany.

Timothy Saunders is a group leader at MBI. His group works on understanding how embryos reliably develop from a single fertilised cell to a complex organism. He has taught biology to adult learners who had previously failed their school science exams (one lesson was published in *Science in School*, see Saunders, 2011) and he is currently involved with the Science outreach and education committee at MBI.

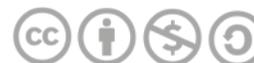

CC BY-NC-SA

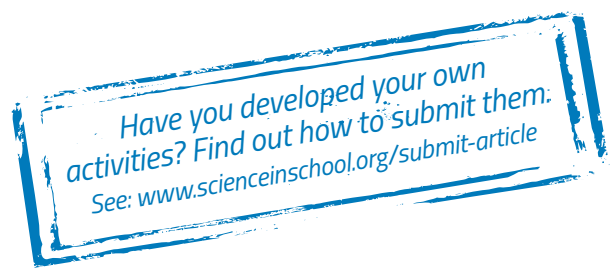

Have you developed your own activities? Find out how to submit them. See: www.scienceinschool.org/submit-article

---

## About *Science in School*

*Science in School* is the only teaching journal to cover all sciences and target the whole of Europe and beyond. The free quarterly journal is printed in English and distributed across Europe. The website is also freely available, offering articles in 30+ languages.

*Science in School* is published and funded by EIROforum (www.eiroforum.org), a partnership between eight of Europe's largest inter-governmental scientific research organisations.

With very few exceptions, articles in *Science in School* are published under Creative Commons licences, so that you can copy and republish the text non-commercially. See www.scienceinschool.org/copyrightViews and opinions expressed by authors and advertisers are not necessarily those of the editors or publisher.

## Advertising: tailored to your needs

For details of how to advertise on the *Science in School* website or in the print journal, see www.scienceinschool.org/advertising or contact advertising@scienceinschool.org